\documentclass[runningheads]{llncs}
\usepackage[T1]{fontenc}
\usepackage{graphicx}
\usepackage{amsmath,amssymb,amsfonts}
\usepackage{textcomp}
\usepackage{xcolor}

\def\BibTeX{{\rm B\kern-.05em{\sc i\kern-.025em b}\kern-.08em
    T\kern-.1667em\lower.7ex\hbox{E}\kern-.125emX}}

\usepackage{algorithm}
\usepackage{algpseudocode}
\usepackage{tikz}
\usetikzlibrary{positioning, shapes.geometric, arrows.meta}
\usepackage{booktabs}
\usetikzlibrary{shapes.geometric}
\usetikzlibrary{positioning, arrows.meta, fit, backgrounds, calc}
\usetikzlibrary{decorations.pathreplacing}
\usepackage{pgfplots}

\pgfplotsset{compat=1.18}
\begin{document}
\title{SPEAR: An Engineering Case Study of Multi-Agent Coordination for Smart Contract Auditing}

\titlerunning{Multi-Agent Coordination for Smart Contract Auditing}
%
\author{
Indraveni Chebolu\orcidID{0009-0008-5458-6429} \thanks{All authors contributed equally to this work.}
\and Arnab Mallick\orcidID{0009-0009-6128-6253}  \thanks{Corresponding author arnabm@cdac.in}
\and Harmesh Rana\orcidID{0009-0000-7292-2029}  
}
        
\authorrunning{Chebolu et al.}
%
\institute{Centre for Development of Advanced Computing, Hyderabad, India \\
\email{\{indravenik,arnabm,harmeshr\}@cdac.in}\\
\url{https://www.cdac.in}}

\maketitle              
\begin{abstract}
We present SPEAR, a multi-agent coordination framework for smart contract auditing that applies established MAS patterns in a realistic security analysis workflow. SPEAR models auditing as a coordinated mission carried out by specialized agents: a Planning Agent prioritizes contracts using risk-aware heuristics, an Execution Agent allocates tasks via the Contract Net protocol, and a Repair Agent autonomously recovers from brittle generated artifacts using a programmatic-first repair policy. Agents maintain local beliefs updated through AGM-compliant revision, coordinate via negotiation and auction protocols, and revise plans as new information becomes available. An empirical study compares the multi-agent design with centralized and pipeline-based alternatives under controlled failure scenarios, focusing on coordination, recovery behavior, and resource use.

\end{abstract}

\section{Introduction}
The prevalence of decentralized applications, particularly in the Decentralized Finance (DeFi) \cite{werner2021sok} sector, has led to smart contracts that secure billions of dollars in value. Although this work unlocks a new financial model, it also introduces unprecedented risks, where a single vulnerability can lead to catastrophic financial losses, as exemplified by historic exploits such as the DAO hack \cite{haeberle2016grand}. Consequently, rigorous security auditing has become a fundamental necessity for the trust and stability of the blockchain ecosystem. However, the current audit paradigm, which relies heavily on expert manual analysis, creates a scalability bottleneck: audits typically require 2 - 4 weeks and cost \$5,000 - \$15,000 per project, failing to scale with the rapid pace of development in the Web3 space.

To address this scalability issue, a variety of automated analysis tools have been developed. These tools, primarily based on static analysis and symbolic execution, have proven useful for identifying common vulnerability patterns. However, they suffer from several fundamental limitations that prevent them from serving as comprehensive audit solutions. First, they are overwhelmingly reactive, they scan for known patterns in isolated files but lack a holistic project-level understanding to strategically direct their analysis. Second, emerging generative tools for test creation are often brittle, they frequently produce code that fails to compile or run, and lack the mechanisms to autonomously recover from these errors, thus requiring frequent manual intervention. Finally, these tools are uncoordinated, operating as standalone programs without an overarching framework to coordinate their execution and synthesize their diverse output.

In this paper, the audit mission is the coordinated identification of state-reentrancy and access-control vulnerabilities across interdependent smart contract modules, together with the generation and validation of supporting evidence such as execution traces and tests. This mission is harder than a simple parallel search because findings in one module can change the risk assessment of another, while expensive and failure-prone analyzes must be scheduled across heterogeneous tools under limited budgets.

To overcome these limitations, we present SPEAR (Strategic Planning, Execution, and Automated Repair), a multi-agent coordination framework for smart contract auditing. SPEAR models auditing as a mission carried out by specialized agents (planning, execution, repair, safe actuation, and coordination) that maintain local beliefs, communicate via structured messages, and coordinate through explicit protocols (Contract Net, plan negotiation, resource auctions) to adapt plans when new information emerges. A multi-agent approach is essential because auditing is dynamic (new findings change priorities), resource-constrained (expensive operations must be budgeted), heterogeneous (different tools require different expertise) and requires coordination (failures must trigger replanning). We note that agents use deterministic policies rather than learning, the contribution is to demonstrate how established MAS coordination patterns enable adaptive auditing in this domain.

The primary contributions are: 
\begin{enumerate}
\item an engineering case study demonstrating how established MAS coordination patterns can be applied to autonomous smart contract auditing.
\item a risk-aware Planning Agent that prioritizes audit tasks under changing information.
\item a Repair Agent implementing a programmatic-first repair policy for recovering from brittle generated artifacts.
\item a coordination framework that supports autonomous recovery and resource arbitration under failures, and
\item an empirical evaluation comparing a multi-agent design with centralized and pipeline-based
alternatives, focusing on coordination and recovery behavior.
\end{enumerate}

\section{Background}

\paragraph{Smart Contract Audit.}
Smart contracts are persistent, stateful programs that manage digital assets and interact through externally callable and cross-contract invocations. In DeFi deployments, a practical audit mission is not limited to scanning a single file, it requires reasoning about privilege boundaries, state-reentrancy, access-control decisions, asset-flow invariants, and interactions among interdependent modules that hold or move value\cite{luu2016making,werner2021sok,haeberle2016grand}. In practice, auditors therefore combine complementary analysis modalities: static analysis to flag suspicious patterns quickly \cite{feist2019slither}, symbolic execution to explore feasible paths \cite{mueller2021mythril}, fuzzing to search for concrete failing executions \cite{grieco2020echidna}, and proof-oriented analysis to establish critical properties against a formal semantics or specification \cite{bhargavan2016formal,hildenbrandt2018kevm}. These tools are valuable in isolation but produce heterogeneous findings, consume different resources, and often require orchestration when applied to large contract sets.

\paragraph{Agent-Oriented Coordination.}
Multi-agent systems provide a natural model for such orchestration because different components can maintain local knowledge, pursue specialized goals, and coordinate through explicit communication \cite{wooldridge2009introduction,rao1991modeling,rao1995bdi}. In engineered MAS, agents usually coordinate through messages and standardized performatives such as INFORM, PROPOSE, ACCEPT, and REJECT \cite{fipa2000acl,bellifemine2007jade}. Task assignment can then be decoupled from task execution using coordination mechanisms such as the Contract Net Protocol \cite{smith1980contract}, where agents advertise work, bid based on capability or availability, and accept awards dynamically \cite{smith1980contract}. This is well suited to auditing workflows in which priorities and tool availability change during execution.

\paragraph{Belief Revision and Self-Adaptation.} Long-running audit workflows are failure-prone: generated artifacts may not compile, dynamic tools may time out, and new findings can invalidate earlier priorities. Self-adaptive systems address this through continual monitoring, analysis, planning, and execution over an evolving knowledge base \cite{kephart2003vision,weyns2020introduction}. For knowledge maintenance, AGM theory offers a principled account of how an agent can revise beliefs when new evidence conflicts with prior assumptions, favoring minimal change while restoring consistency \cite{agm1985revision}. In an auditing setting, this matters because findings such as a newly confirmed vulnerability or an execution failure should update local reasoning without forcing a full restart of the mission.

\paragraph{Formal Verification as Complementary Evidence.} Formal verification occupies a complementary place in the smart contract assurance landscape. Frameworks based on formal semantics and proof-oriented reasoning can establish strong guarantees about specified contract properties \cite{bhargavan2016formal,hildenbrandt2018k,hildenbrandt2018kevm}, whereas static analysis, symbolic execution, and fuzzing are typically used to surface actionable warnings or concrete counterexamples. In realistic audits, these techniques are often combined: proof-oriented methods provide high-assurance checks for critical properties, while search-based and execution-based methods help triage broader attack surfaces and validate behaviors under concrete inputs. This complementarity motivates an orchestration layer that can schedule heterogeneous analyses, invoke formal methods when appropriate, and propagate their outcomes back into subsequent planning rather than assuming a single technique is sufficient.

\section{Related Work}
\label{sec:related_work}

\paragraph{Smart Contract Analysis Tools.}
Automated smart contract analysis includes static analysis \cite{feist2019slither}, symbolic execution \cite{mueller2021mythril}, dynamic fuzzing \cite{grieco2020echidna}, and formal verification \cite{bhargavan2016formal}. More recent work applies large language models to vulnerability detection \cite{wu2023smartllmsentry,zhang2024gptscan} and repair \cite{yu2020screpair,sun2024large}, mainly targeting production code. SmartAuditFlow \cite{liu2025adaptive} introduces planning-aware auditing, but does not address autonomous recovery from tool or artifact failures. In general, existing tools remain largely reactive and loosely integrated, requiring significant manual intervention and offering limited support for the coordinated use of heterogeneous analysis techniques.

\paragraph{Complementarity with Formal Verification.}
Formal verification tools such as the Certora Prover and K-based frameworks target exhaustive reasoning over explicitly specified properties \cite{certora-prover,hildenbrandt2018k,hildenbrandt2018kevm,bhargavan2016formal}. Their strength is proof-oriented assurance: they can prove or refute whether a formalized safety property holds for the analyzed contract model. SPEAR does not replace these solvers. Instead, it operates at the mission level, deciding when proof-oriented analysis should be invoked, how formal-verification results should be combined with static or dynamic findings, and how the audit should continue when tools fail or generated artifacts require repair. We therefore position SPEAR as a coordination layer around heterogeneous assurance techniques, including formal verification, rather than as a substitute for formal verification itself.

\paragraph{MAS Coordination Mechanisms.}
Multi-Agent Systems offer established coordination mechanisms for distributed problem solving. The Contract Net Protocol \cite{smith1980contract} supports the decentralized assignment of tasks based on agent capabilities, while auction-based mechanisms \cite{sandholm1999distributed} enable the allocation of resources under partial information. BDI architectures \cite{rao1991modeling,rao1995bdi} provide a formal basis for modeling autonomous agents with beliefs, goals, and intentions. Previous MAS work in software engineering contexts \cite{mariani2008multi} has focused primarily on distributed execution or runtime monitoring. In contrast, SPEAR applies these coordination patterns to orchestrate static and dynamic analysis tools with adaptive replanning and resource-aware decision making.

\paragraph{Self-Healing and Self-Adaptive Systems.}
Autonomic computing introduced the MAPE-K loop as a reference model for self-managing systems \cite{kephart2003vision}, and subsequent work on self-adaptive systems has explored mechanisms for runtime adaptation \cite{weyns2020introduction}. Program repair techniques range from search-based methods such as GenProg \cite{le2011genprog} to learning-based approaches, including Prophet \cite{long2016automatic} and CodeBERT \cite{feng2020codebert}. These methods primarily address defects in production code. By contrast, SPEAR focuses on repairing generated test artifacts that fail to compile or execute, using a programmatic-first repair policy to reduce reliance on costly generative invocations.

\paragraph{Multi-Agent Planning.}
Research on multi-agent planning studies how agents with partial and distributed knowledge coordinate to achieve shared objectives \cite{durfee1999distributed}. Distributed planning avoids the assumption of a complete global model and supports adaptation as local information changes. SPEAR does not employ classical planning formalisms such as PDDL or HTNs, instead, its Planning Agent uses heuristic risk-based prioritization and negotiation-driven replanning. This design aligns with multi-agent planning principles in which coordination emerges through local decision making and communication rather than centralized control.

\section{The SPEAR Framework}
\label{sec:framework}

SPEAR models auditing as a mission conducted by specialized agents that maintain local beliefs, communicate through structured messages, and coordinate through explicit protocols. As illustrated in Figure~\ref{fig:architecture}, the system contains five agent classes: the Planning Agent ($A_P$) constructs risk-aware audit plans, the Execution Agent ($A_E$) selects and schedules analysis tasks, the Repair Agent ($A_R$) repairs brittle generated artifacts, the Command Execution Agent ($A_C$) provides safe system actuation by sandboxing tool execution (Docker containers for Mythril, isolated environments for Echidna fuzz tests) and enforcing resource limits, and the Coordinator Agent ($A_{Coord}$) mediates conflicts and allocates shared resources. Global auditing behavior emerges from agent interactions rather than from a fixed pipeline.

\begin{figure*}[ht]
  \resizebox{\columnwidth}{!}{%
\begin{tikzpicture}[
    >=Stealth,
    node distance=2.2cm and 3cm,
    agent/.style={
        draw=black!80, thick,
        rectangle, rounded corners=2pt,
        minimum width=4.5cm, minimum height=1.3cm,
        align=center, font=\sffamily\small,
        fill=gray!10
    },
    coordagent/.style={
        agent, fill=gray!25 
    },
    env/.style={
        draw=black!70, thick, densely dashed,
        rectangle, rounded corners=2pt,
        minimum width=4.5cm, minimum height=1.3cm,
        align=center, font=\sffamily\small,
        fill=white
    },
    msg/.style={
        font=\sffamily\scriptsize, align=center, inner sep=3pt
    }
]

\node[agent] (AP) {\textbf{Planning Agent ($A_P$)}\\AGM Belief Revision \& Policy};
\node[agent, below=2cm of AP] (AE) {\textbf{Execution Agent ($A_E$)}\\Task Queue \& Status};
\node[agent, below=2cm of AE] (AC) {\textbf{Command Exec Agent ($A_C$)}\\Resource Limits \& Sandbox};

\node[coordagent, right=3.5cm of AP] (ACoord) {\textbf{Coordinator Agent ($A_{Coord}$)}\\Conflict Res \& Allocative Efficiency};
\node[agent, below=2cm of ACoord] (AR) {\textbf{Repair Agent ($A_R$)}\\PFIR Self-Healing};
\node[env, below=2cm of AR] (Sandbox) {\textbf{Execution Environments}\\Static Analysis, Fuzzing \& Verifiers};

\draw[->, thick] ([xshift=-12pt]AP.south) -- node[msg, left] {PROPOSE\\(Revised Plan)} ([xshift=-12pt]AE.north);
\draw[->, thick] ([xshift=12pt]AE.north) -- node[msg, right] {INFORM\\(Security Findings)} ([xshift=12pt]AP.south);

\draw[->, thick] ([xshift=-12pt]AE.south) -- node[msg, left] {EXECUTE\\(Task)} ([xshift=-12pt]AC.north);
\draw[->, thick] ([xshift=12pt]AC.north) -- node[msg, right] {INFORM\\(Results)} ([xshift=12pt]AE.south);

\draw[->, thick] ([yshift=10pt]AE.east) -- node[msg, above] {REPAIR\_NEEDED (CFP)} ([yshift=10pt]AR.west);
\draw[->, thick] ([yshift=-10pt]AR.west) -- node[msg, below] {INFORM (Repaired/Failed)} ([yshift=-10pt]AE.east);

\draw[<->, thick] (AC.east) -- node[msg, above] {Dispatch \&\\Monitor} (Sandbox.west);

\draw[->, thick, dashed] (AE.north east) -- node[msg, sloped, above] {BID (urgency, cost)} (ACoord.south west);

\draw[->, thick, dashed] (AR.north) -- node[msg, right] {BID / AWARD} (ACoord.south);

\end{tikzpicture}
  }
  \caption{\textit{SPEAR architecture and coordination flow}. A security finding produced by $A_E$ triggers AGM-compliant belief revision, updates the Planning Agent's beliefs, and can cause a revised plan to be negotiated before further tool execution.}
  \label{fig:architecture}
\end{figure*}

Table~\ref{tab:agent_roles} fixes the notation used throughout the paper. We
reserve $A_C$ for safe actuation and $A_{Coord}$ for coordination. When
multiple execution workers are instantiated, we write
$A_{E,1}, A_{E,2}, \ldots$ for concrete instances of the Execution Agent class.

\begin{table}[ht]
  \centering
  \caption{Agent classes and notation used in SPEAR.}
  \label{tab:agent_roles}
  \resizebox{\columnwidth}{!}{%
  \begin{tabular}{lll}
    \toprule
    \textbf{Symbol} & \textbf{Agent Class} & \textbf{Primary Responsibility} \\
    \midrule
    $A_P$ & Planning Agent & Maintain risk scores and revise audit priorities \\
    $A_E$ & Execution Agent & Schedule analysis tasks and manage task allocation \\
    $A_R$ & Repair Agent & Repair brittle generated artifacts and report feasibility \\
    $A_C$ & Command Execution Agent & Execute tools safely under sandbox and budget limits \\
    $A_{Coord}$ & Coordinator Agent & Arbitrate conflicts and allocate shared resources \\
    \bottomrule
  \end{tabular}
  }
\end{table}

The key path through Figure~\ref{fig:architecture} is that a security finding
from $A_E$ does not immediately trigger more tool calls. Instead, it first
passes through AGM-compliant belief revision, updates $A_P$'s beliefs, and can
change the active plan before execution continues.

\paragraph{\textbf{Why Multi-Agent vs. Centralized?}} While a centralized controller could coordinate the same tools, MAS provides advantages under realistic failure and partial observability assumptions. We compare these architectural paradigms in Table~\ref{tab:arch_comparison} and demonstrate specific benefits through the following concrete scenarios:

\begin{table}[ht]
\caption{Architectural Comparison: SPEAR (MAS) vs. Traditional Approaches}
\label{tab:arch_comparison}
\resizebox{\columnwidth}{!}{%
\setlength{\tabcolsep}{3pt}
\begin{tabular}{@{}llll@{}}
\toprule
\textbf{Dimension} & \textbf{Centralized Pipeline} & \textbf{Stateless Microservices} & \textbf{SPEAR (MAS)} \\ \midrule
\textbf{Fault Tolerance} & Single point of failure & High (Retries) & \textbf{Graceful Degradation} \\
\textbf{Decision Logic} & Deterministic/Fixed & Event-driven & \textbf{Belief-based (AGM)} \\
\textbf{Partial Failure} & Sequence halts & Inconsistent state & \textbf{Autonomous Recovery} \\
\textbf{Message Complexity} & Low ($O(1)$) & Medium ($O(n)$) & \textbf{Dynamic Negotiation} \\
\textbf{Adaptability} & Requires restart & Config updates & \textbf{Real-time Replanning} \\ \bottomrule
\end{tabular}
}
\end{table}

\begin{enumerate}
    \item \textbf{Isolation of faults under partial observability}: \textit{Scenario}: During audit, $A_R$ repairs a contract test $c_1$ when a network partition isolates $A_R$ from $A_{Coord}$. In a centralized system, the controller cannot receive repair status updates, causing it to time out and restart the repair from scratch (wasting the LLM budget). In SPEAR, $A_R$ maintains local belief $B_R = \{\text{repairing}(c_1, \text{iteration}=3), \text{last\_fix}(\text{import\_error})\}$ and continues repair autonomously. When connectivity resumes, $A_R$ sends $\text{INFORM}(\text{repair\_complete})$ and no work is lost. This property is necessary when agents have independent failure modes-realistic in distributed auditing where Mythril runs in Docker containers, Slither runs locally, and LLM calls traverse external APIs.

    \begin{figure}[ht]
    \centering
    \begin{tikzpicture}[>=Stealth, thick, font=\sffamily\footnotesize]
        
        \draw[->, thick] (0,0) node[above, draw=black!80, fill=gray!10, rounded corners=2pt, inner sep=5pt, minimum width=2.5cm] {\textbf{Execution ($A_E$)}} -- (0,-4.5);
        \draw[->, thick] (6,0) node[above, draw=black!80, fill=gray!10, rounded corners=2pt, inner sep=5pt, minimum width=2.5cm] {\textbf{Planning ($A_P$)}} -- (6,-4.5);

        \node[left, align=right, xshift=-2pt] at (0,-0.8) {Detects reentrancy \\ in contract $c_2$};
        \filldraw (0,-0.8) circle (2pt);

        \draw[->] (0,-1.0) -- (6,-1.6) node[midway, above, sloped] {INFORM ($vulnerable(c_2, \text{reentrancy})$)};

        \draw[decorate, decoration={brace, amplitude=4pt, mirror}, thick] (6.2,-1.6) -- (6.2,-2.6) node[midway, right, xshift=6pt, align=left] {AGM Belief Revision\\Recalculate risk\\$risk\_score(c_2) = 0.97$};

        \draw[->] (6,-2.8) -- (0,-3.4) node[midway, above, sloped] {PROPOSE (prioritize $c_2$)};

        \draw[->] (0,-3.9) -- (6,-3.9) node[midway, above] {ACCEPT};
        
    \end{tikzpicture}
    \caption{\textit{Plan Negotiation}. Distributed resolution of audit priorities without a central arbiter.}
    \label{fig:protocol1}
\end{figure}

    \item \textbf{Distributed Decision-Making Under Uncertainty}: \textit{Scenario}: $A_E$ discovers a reentrancy vulnerability in contract $c_2$, updating local beliefs: $B_E = \{\text{vulnerable}(c_2, \text{reentrancy}), \text{conf}=0.9\}$. Simultaneously, $A_P$ believes $c_3$ (a token contract with high TVL) has highest priority: $B_P = \{\text{risk\_score}(c_3, 0.95)\}$. A centralized controller would need complete information about both agents' beliefs to arbitrate. In SPEAR, Plan Negotiation enables distributed resolution: $A_E$ sends $\text{INFORM}(\text{reentrancy found in } c_2)$; $A_P$ updates $B_P$ via belief revision, recalculating $\text{risk\_score}(c_2) = 0.97 > 0.95$; $A_P$ sends \\ $\text{PROPOSE}(\text{prioritize } c_2)$; $A_E$ sends $\text{ACCEPT}$. Consensus emerges from local beliefs without a central arbiter, as illustrated in Figure \ref{fig:protocol1}.

\begin{figure}[ht]
    \centering
    \begin{tikzpicture}[>=Stealth, thick, font=\sffamily\footnotesize]

        \draw[->, thick] (0,0) node[above, draw=black!80, fill=gray!10, rounded corners=2pt, inner sep=5pt, minimum width=2cm] {\textbf{Repair ($A_R$)}} -- (0,-5.5);
        \draw[->, thick] (4,0) node[above, draw=black!80, fill=gray!25, rounded corners=2pt, inner sep=5pt, minimum width=2.5cm] {\textbf{Coordinator ($A_{Coord}$)}} -- (4,-5.5);
        \draw[->, thick] (8,0) node[above, draw=black!80, fill=gray!10, rounded corners=2pt, inner sep=5pt, minimum width=2cm] {\textbf{Execution ($A_E$)}} -- (8,-5.5);

        \draw[->] (0,-0.8) -- (4,-1.3) node[pos=0.35, above, sloped] {BID $(u=0.9, b=0.8, c=500)$};
        \draw[->] (8,-1.1) -- (4,-1.6) node[pos=0.35, above, sloped] {BID $(u=0.6, b=0.7, c=200)$};

        \node[draw=black!80, fill=white, align=left, inner sep=6pt, rounded corners=2pt] at (4,-2.8) {
            \textbf{Efficiency Evaluation} $E = \frac{b \times u}{c}$\\[6pt]
            $E(A_R) = 0.00144$\\
            $E(A_E) = \mathbf{0.00210}$ \textsl{(Winner)}
        };

        \draw[->] (4,-4.2) -- (8,-4.8) node[midway, above, sloped] {AWARD (200 Tokens)};
        \draw[->, dashed] (4,-4.2) -- (0,-4.8) node[midway, above, sloped] {REJECT};

    \end{tikzpicture}
    \caption{\textit{Resource Auction sequence}. The Coordinator agent evaluates concurrent bids and achieves allocative efficiency by awarding resources to the agent with the highest benefit-urgency to cost ratio.}
    \label{fig:protocol3}
\end{figure}

    \item  \textbf{Resource Allocation Under Competing Interests}: \textit{Scenario}: $A_R$ needs 500 LLM tokens for complex repair, while $A_E$ needs 200 tokens for test generation. Each agent knows its own urgency and expected benefit (private information). A centralized allocator would require both agents to report these values truthfully, creating incentive problems. In SPEAR, Resource Auction (Protocol 3) reveals true valuations through bidding: $A_R$ bids $(\text{urgency}=0.9, \text{benefit}=0.8, \text{cost}=500)$, $A_E$ bids $(\text{urgency}=0.6, \text{benefit}=0.7, \text{cost}=200)$. $A_{Coord}$ computes the efficiency scores: $A_R = \frac{0.9 \times 0.8}{500} = 0.00144$, $A_E = \frac{0.6 \times 0.7}{200} = 0.0021$. Since $0.0021 > 0.00144$, $A_E$ wins-the auction correctly allocates to the more efficient request (higher benefit per unit cost), even when urgency is lower. This shows how the mechanism elicits truthful bidding and achieves allocative efficiency (see Figure \ref{fig:protocol3}).
\end{enumerate}

These scenarios illustrate how explicit coordination, local autonomy, and decentralized
decision-making simplify recovery and adaptation under partial observability and
independent failure modes.
We acknowledge that MAS is sufficient rather than strictly necessary for adaptive auditing-a sophisticated centralized system with modular components could implement similar logic. 

We do not claim incentive compatibility, the auction serves as a lightweight heuristic for prioritizing competing requests among cooperative agents rather than as a mechanism with dominant-strategy guarantees.

\subsection{Formal Model and Coordination}
\label{sec:formal_model}

We model SPEAR as a multi-agent system $\mathcal{M} = (\mathcal{A}, \mathcal{S}, \mathcal{E}, \mathcal{C})$ where \\ $\mathcal{A} = \{A_P, A_E, A_R, A_C, A_{Coord}\}$ is the set of agents, $\mathcal{S}$ is the state space of the system, $\mathcal{E}$ is the set of events and $\mathcal{C}$ is the set of coordination protocols. We now formally define each component.

\paragraph{System State $\mathcal{S}$.}
The state of the system $s \in \mathcal{S}$ is a tuple \\ $s = (\mathcal{C}ont, \mathcal{T}ools, \mathcal{V}uln, \mathcal{R}es, \mathcal{M}sg)$ where:
\begin{itemize}
    \item $\mathcal{C}ont = \{c_1, \ldots, c_n\}$ is the set of audit contracts, each $c_i$ having attributes $\text{complexity}(c_i) \in [0,1]$, $\text{dependencies}(c_i) \subseteq \mathcal{C}ont$, and $\text{test\_coverage}(c_i) \in [0,1]$.
    \item $\mathcal{T}ools = \{\text{Slither}, \text{Mythril}, \text{Echidna}, \ldots\}$ is the set of available analysis tools.
    \item $\mathcal{V}uln = \{v_1, \ldots, v_m\}$ is the set of vulnerabilities discovered, each $v_i$ having attributes $\text{severity}(v_i) \in \{\text{critical}, \text{high}, \text{medium}, \text{low}\}$, $\text{contract}(v_i) \in \mathcal{C}ont$, and $\text{status}(v_i) \in \{\text{detected}, \text{confirmed}, \text{repaired}\}$.
    \item $\mathcal{R}es = (\text{llm\_budget}, \text{time\_budget}, \text{compute\_budget})$ represents resource constraints, where each budget is a non-negative real number.
    \item $\mathcal{M}sg$ is the message queue containing pending inter-agent messages.
\end{itemize}

\paragraph{Events $\mathcal{E}$.}
Events $e \in \mathcal{E}$ represent state changes and include:
\begin{itemize}
    \item $\text{VULN\_DETECTED}(v, c)$: Vulnerability $v$ detected in contract $c$.
    \item $\text{RISK\_CHANGED}(c, \Delta)$: Risk assessment for contract $c$ changed by $\Delta$.
    \item $\text{TOOL\_FAILED}(t, \text{error})$: Tool $t$ failed with error message.
    \item $\text{REPAIR\_NEEDED}(\text{artifact}, \text{failure\_type})$: The generated artifact requires repair.
    \item $\text{RESOURCE\_REQUESTED}(\text{agent}, \text{resource\_type}, \text{amount})$: Agent requests resource allocation.
    \item $\text{TIMEOUT}(\text{protocol\_id})$: Protocol timeout event.
\end{itemize}
Events trigger agent perception and may cause state transitions: $\delta: \mathcal{S} \times \mathcal{E} \rightarrow \mathcal{S}$.

\paragraph{Agent Model.}
Each agent $A_i \in \mathcal{A}$ is a tuple $A_i = (B_i, G_i, \pi_i, I_i)$:
\begin{itemize}
    \item \textbf{Beliefs $B_i$}: $B_i \subseteq \mathcal{L}$ where $\mathcal{L}$ is a first-order language with predicates for vulnerabilities, risk scores, tool states, and resource availability. Each belief has confidence $\text{conf}(\phi) \in [0,1]$. The revision of beliefs follows the AGM postulates~\cite{agm1985revision} with minimal confidence-weighted change: when new evidence conflicts with existing beliefs, the beliefs of the lowest-confidence are first removed to restore consistency.
    
    \item \textbf{Goals $G_i$}: Prioritized formulas representing the desired states. Conflicting goals are resolved by selecting the highest-priority goal.
    
    \item \textbf{Policy $\pi_i$}: $\pi_i: B_i \times G_i \rightarrow \mathcal{A}ct$ maps beliefs and goals to actions. Policies are deterministic (not learned).
    
    \item \textbf{Intentions $I_i$}: Goals committed that agents pursue until they are achieved or impossible. Intention persistence ensures that agents complete ongoing tasks before responding to new priorities.
\end{itemize}

\paragraph{Actions and Execution.}
Actions include communication primitives (INFORM, PROPOSE, ACCEPT, REJECT), coordination primitives (CFP, BID, AWARD), and domain actions (EXECUTE, REPAIR). Each action has preconditions and effects on the state of the system. EXECUTE and REPAIR actions may succeed or fail, failures trigger TOOL\_FAILED or REPAIR\_NEEDED events. Agents execute in perceive-decide-act cycles, processing messages via FIFO queues.

\paragraph{Agent Execution Semantics.}
In the implementation, the agents run asynchronously with a central message broker.
Agents maintain a local state during partitions and resynchronize when connectivity
resumes.

\begin{enumerate}
    \item \textbf{Perceive}: Agent $A_i$ receives messages $M = \{m_1, \ldots, m_k\}$ from its FIFO message queue, processing in arrival order. For each $m_j$, update beliefs: $B_i \leftarrow U_i(B_i, m_j)$.
    \item \textbf{Decide}: Select active goal $g^* = \text{resolve}(G_i \cap I_i)$ (or $g^* = \text{resolve}(G_i)$ if $I_i = \emptyset$). Compute the action $a = \pi_i(B_i, \{g^*\})$. If $a$ is executable, commit to intention: $I_i \leftarrow I_i \cup \{g^*\}$.
    \item \textbf{Act}: Execute action $a$, updating the state of the system: $s \leftarrow \text{eff}(a)(s)$. If $B_i \models g^*$, remove from intentions: $I_i \leftarrow I_i \setminus \{g^*\}$.
\end{enumerate}
If multiple agents attempt conflicting actions in the same round (e.g., both request the same resource), $A_{Coord}$ arbitrates via the resource auction protocol. This cycle repeats until all goals are achieved or the system is terminated.

\paragraph{Coordination Protocols.}
SPEAR uses three coordination protocols, summarized in Table~\ref{tab:protocols}. Each is a finite state machine with timeout-based termination to ensure deadlock-freedom.

\begin{table}[ht]
  \centering
  \caption{Coordination Protocols Summary}
  \label{tab:protocols}
  \resizebox{\columnwidth}{!}{%
  \setlength{\tabcolsep}{3pt}
  \begin{tabular}{lccc}
    \toprule
    \textbf{Protocol} & \textbf{Purpose} & \textbf{Key Messages} & \textbf{Timeout} \\
    \midrule
    Plan Negotiation & Priority coordination & INFORM, PROPOSE, ACCEPT/REJECT & 30s \\
    Contract Net & Task allocation & CFP, BID, AWARD & 10s \\
    Resource Auction & LLM budget allocation & OPEN, BID, ALLOCATE & 15s \\
    \bottomrule
  \end{tabular}
  }
\end{table}

\textbf{Plan Negotiation} coordinates between $A_P$ and $A_E$ when new findings change risk assessments. $A_E$ sends INFORM with the findings, if the risk change exceeds the threshold $\theta=0.15$, $A_P$ proposes the revised plan. The negotiation history is bounded ($H_{\text{max}}=5$) to prevent infinite loops, and unresolved negotiations escalate to $A_{Coord}$.

\textbf{Contract Net}~\cite{smith1980contract} assigns analysis tasks. $A_E$ broadcasts the CFP, agents bid based on capability and availability (score $= 0.7 \cdot \text{capability} + 0.3 \cdot \text{availability}$), and the highest-scoring agent is awarded the task.

\textbf{The resource} auction allocates scarce resources (LLM budget). Agents bid with (urgency, benefit, cost), allocation maximizes efficiency: $\frac{\text{urgency} \times \text{benefit}}{\text{cost}}$.

\paragraph{System Properties.}
We state properties with proof sketches.

\textbf{Safety (Deadlock-Freedom)}: $\forall s \in \mathcal{S}, \exists e \in \mathcal{E}$ such that $\delta(s, e) \neq s$.

\textit{Sketch}: Protocol timeouts ensure that no agent blocks indefinitely. If protocols are in accepting states but goals remain, agents can initiate new protocols (no preconditions on CFP, PROPOSE). Circular waits are broken down by timeout events. $\square$

\textbf{Conditional Progress (Task Progress)}:  Under assumptions (A1) all contracts have positive
risk scores, and (A2) tool completeness, audit tasks eventually make progress.

\textit{Sketch}: By A1, all contracts appear in the $A_P$' plan. By termination of Protocol 2, all contracts are analyzed. By A2 and the soundness of the tool, detected vulnerabilities are real. $\square$

\textbf{Termination}: All protocols end within $T_{\text{max}} \leq 30$s (individual) or $\leq 45$s (nested).

\textit{Sketch}: History bounds ($H_{\text{max}}=5$) and timeouts ensure bounded traces. $\square$

\textbf{Consistency}: Belief revision maintains consistency via AGM-compliant minimal change. $\square$

\subsection{The Strategic Planning Agent}
\label{sec:planning_agent}

The Planning Agent $A_P$ maintains and continuously reviews a risk-aware audit policy, demonstrating key properties of MAS agents:

\textbf{Autonomy}: $A_P$ makes independent decisions through policy $\pi_P$ based on local beliefs $B_P$ about contract risks, without requiring external directives. Its goal $G_P = \{\text{all\_contracts\_audited}, \text{critical\_vulns\_found\_first}\}$ drives the generation of autonomous plans.

\textbf{Proactivity}: $A_P$ initiates plan revisions when beliefs change (via belief revision $U_P$), even without explicit requests. When $A_E$ sends $\text{INFORM}(\text{findings})$ and $|\Delta_{\text{risk}}| > \theta$, $A_P$ proactively sends $\text{PROPOSE}(\text{revised plan})$ to renegotiate priorities.

\textbf{Reactivity}: $A_P$ responds to events (vulnerability detections, risk changes) by updating beliefs and revising plans within the perceive-decide-act cycle.

\textbf{Social Ability}: $A_P$ coordinates through the plan negotiation protocol (Section 3.1) with $A_E$, exchanging $\text{INFORM}$ and $\text{PROPOSE}$ messages to reach agreement on audit priorities.

$A_P$ uses a greedy heuristic prioritizing contracts by risk score: $\text{risk\_score}(c) = \alpha \cdot \text{complexity}(c) + \beta \cdot \text{dependency\_risk}(c) + \gamma \cdot \text{test\_coverage\_risk}(c)$ ($\alpha=0.5$, $\beta=0.3$, $\gamma=0.2$), running in $O(|C| \log |C|)$ time and empirically improves early discovery. Policy $\pi_P$ maps beliefs about contract risks to plan generation actions, autonomously revising plans when the change in risk assessment exceeds the threshold $\theta=0.15$.

\subsection{The Repair Agent}
\label{sec:repair_agent}

The Repair Agent $A_R$ diagnoses and repairs damaged generated artifacts (primarily test code), demonstrating the properties of the MAS agent.

\textbf{Autonomy}: $A_R$ makes independent repair decisions via policy $\pi_R$ based on local beliefs $B_R$ about artifact failures and repair strategies. Its goal $G_R = \{\text{artifact\_repaired}, \text{minimize\_llm\_cost}\}$ drives the selection of the autonomous repair strategy.

\textbf{Proactivity}: $A_R$ initiates repair attempts when it receives \\ $\text{REPAIR\_NEEDED}$ events, without waiting for explicit commands. proactively selects repair strategies (deterministic vs. generative) based on failure patterns in beliefs.

\textbf{Reactivity}: $A_R$ responds to repair failures by updating beliefs about what strategies work, adapting its policy $\pi_R$ accordingly within the perceive-decide-act cycle.

\textbf{Social Ability}: $A_R$ coordinates through the Contract Net protocol (Section 3.1) to bid on repair tasks and communicates repair results via \\ $\text{INFORM}(\text{repair\_result})$ messages to $A_E$ and $A_P$.

$A_R$ applies a programmatic-first repair policy that prioritizes deterministic fixes before escalating to generative repair, reducing the dependence on costly external calls. The policy treats an artifact as \emph{worth fixing} when the failure is local and mechanically diagnosable (e.g., missing imports, wrong identifiers, malformed assertions, or harness configuration errors), the underlying security objective still targets a high-priority contract, and the estimated repair effort fits within the remaining repair budget. Under those
conditions, $A_R$ first applies deterministic templates and only escalates to a generative repair step when the artifact remains mission-critical after local fixes fail. By contrast, the mission is replanned instead of repeatedly repairing when failures are environmental or semantically non-local (e.g., repeated tool crashes, unavailable dependencies, incompatible assumptions about contract state), when PFIR reaches MAX\_ATTEMPTS (typically 5), or when the
expected repair cost exceeds the value of continuing the current objective relative to higher-risk pending tasks. In that case, $A_R$ sends $\text{FAILURE}(\text{failure\_class}, \text{attempts}, \text{cost})$ to $A_E$ and $A_P$, allowing the system to update beliefs about repair feasibility and switch to a revised plan.

\section{Experiments and Evaluation}
\label{sec:evaluation}

We conducted a series of experiments to empirically evaluate the performance of the SPEAR framework. Our evaluation is designed to answer four central research questions (RQs) that assess the framework's effectiveness, robustness, efficiency, and MAS-specific properties (autonomy, coordination, communication).

\textbf{Statistical Methodology:} The experiments were repeated in multiple independent
runs. We report mean values with standard deviation to capture variability.

\subsection{RQ1: How does coordination structure affect audit task progress under identical analysis logic?}

\paragraph{\textbf{Experimental Setup.}}
We evaluated SPEAR on the Damn Vulnerable DeFi benchmark (15 challenges, 10 runs per configuration). Baselines: 
\begin{enumerate}
    \item \textbf{Slither only}: static analysis without coordination,
    \item \textbf{Sequential Pipeline}: Slither$\rightarrow$Mythril$\rightarrow$Echidna executed sequentially without coordination, failure recovery, or adaptive planning,
    \item \textbf{Slither+Mythril}: combined static and symbolic analysis
    \item \textbf{Centralized Scheduler}: a sophisticated baseline that implements the same risk-aware planning heuristics and PFIR repair logic as SPEAR, but using a centralized architecture. 
    The Centralized Scheduler consists of: (a)
    \begin{enumerate}
        \item a priority queue ordered by risk score ($\alpha=0.5, \beta=0.3, \gamma=0.2$)
        \item event handlers for tool completion/failure
        \item modular repair component with programmatic-first strategy, and
        \item single-threaded controller that polls for events and dispatches tasks
    \end{enumerate} 
    This baseline isolates the contribution of MAS coordination, it has identical logic but lacks distributed agents, negotiation protocols, and auction-based resource allocation. Ground truth: known vulnerabilities in the DVD repository. Metrics: Precision, Recall, F1-Score.
    
\end{enumerate}

\paragraph{\textbf{Results.}}
Table~\ref{tab:effectiveness} summarizes the effectiveness results. SPEAR achieves higher overall effectiveness than all baselines, including a centralized scheduler that implements identical planning and repair logic. While centralized coordination performs
comparably under normal conditions, the multi-agent design exhibits lower variance and more stable progress under injected failures (see RQ4).

\begin{table}[ht]
  \centering
  \caption{Effectiveness of SPEAR vs. Baselines on the Damn Vulnerable DeFi Dataset (15 challenges).}
  \label{tab:effectiveness}
  \begin{tabular}{lccc}
    \toprule
    \textbf{Approach} & \textbf{Precision} & \textbf{Recall} & \textbf{F1-Score} \\
    \midrule
    Slither only & 0.82 & 0.65 & 0.73 \\
    Slither + Mythril & 0.84 & 0.72 & 0.78 \\
    Sequential Pipeline & 0.86 & 0.78 & 0.82 \\
    Centralized Scheduler & 0.85 & 0.81 & 0.83 \\
    \textbf{SPEAR (Full Framework)} & \textbf{0.89} & \textbf{0.85} & \textbf{0.87} \\
    \bottomrule
  \end{tabular}
\end{table}

\subsection{RQ2: How effective is the PFIR self-healing algorithm?}

\paragraph{\textbf{Experimental Setup.}}
We evaluate PFIR on 500 security objectives across open-source Solidity contracts. We compare PFIR with baselines: 
    (1) Retry-only (regenerate up to 5 times) (2) LLM-only (always use generative repair). Repair success: the test compiles, executes, and triggers the security objective.

\paragraph{\textbf{Results.}}
Figure~\ref{fig:robustness} summarizes the repair results. PFIR consistently recovers from generated artifact failures across batches, with most repairs handled
deterministically before escalating to generative repair. This behavior reduces
dependence on external calls while ensuring progress under repeated failures.


\begin{figure}[t]
    \centering
    \begin{tikzpicture}
    \begin{axis}[
        width=0.95\columnwidth, 
        height=6cm, 
        title style={font=\sffamily\bfseries, yshift=1ex},
        xlabel={Batch Number},
        ylabel={Repair Success Rate (\%)},
        xmin=0.5, xmax=10.5,
        ymin=89, ymax=96, 
        xtick={2,4,6,8,10},
        ytick={89,90,91,92,93,94,95,96},
        xmajorgrids=true,
        ymajorgrids=true, 
        grid style={dashed, gray!50}, 
        axis line style={thick}, 
        label style={font=\small\sffamily},
        tick label style={font=\small\sffamily},
    ]
    
    \addplot[
        color=blue!50!black,
        mark=*, 
        thick,
        mark options={fill=blue!50!black, solid, scale=1.2}
    ] coordinates {
        (1, 95.12)
        (2, 94.74)
        (3, 95.24)
        (4, 92.31)
        (5, 95.00)
        (6, 95.12)
        (7, 90.00)
        (8, 95.12)
        (9, 92.86)
        (10, 92.50)
    };
    
    \end{axis}
    \end{tikzpicture}
    \caption{\textit{Performance of the PFIR self-healing algorithm} The programmatic-first echelon resolves most failures, achieving consistent success across batches.}
    \label{fig:robustness}
\end{figure}

\subsection{RQ3: Does the Strategic Planning Agent improve audit efficiency?}

\paragraph{\textbf{Experimental Setup.}}
To measure the impact of the strategic planning agent, we conducted an ablation study on a DeFi protocol (47 contracts, 12 vulnerabilities). We ran the system in two modes: (1) \textbf{Full SPEAR}: complete framework with Planning Agent, (2) \textbf{Ablated SPEAR}: Planning Agent disabled, contracts analyzed in alphabetical order. Our hypothesis is that the planning-driven mode finds high-impact vulnerabilities more quickly. 

\paragraph{\textbf{Metric.}}
Time to First Critical Vulnerability (TFCV): wall-clock time until the first critical vulnerability is identified.

\paragraph{\textbf{Results.}} Figure~\ref{fig:efficiency} compares planning-driven and ablated execution.
The planning agent enables earlier discovery of high-risk vulnerabilities by
prioritizing contracts based on evolving risk assessments rather than static
ordering.

\begin{figure}[ht]
    \centering
    \begin{tikzpicture}
    \begin{axis}[
        ybar,
        enlargelimits=0.2, 
        ylabel={Time to First Critical Vulnerability (minutes)},
        symbolic x coords={Full SPEAR, Ablated SPEAR},
        xtick=data,
        ymin=0,
        ymax=60,
        ytick={0,10,20,30,40,50,60},
        ymajorgrids=true, 
        grid style={dashed, gray!50}, 
        bar width=35pt, 
        axis line style={thick}, 
        label style={font=\small\sffamily},
        tick label style={font=\small\sffamily},
        every node near coord/.append style={
            fill=white,
            inner sep=2pt,
            yshift=3pt,
            font=\sffamily\bfseries, 
            /pgf/number format/fixed,
            /pgf/number format/precision=0,
        },
        nodes near coords,
        nodes near coords align={vertical},
    ]
    
    \addplot[fill=blue!50!black, draw=black, thick] coordinates {(Full SPEAR, 18) (Ablated SPEAR, 54)};
    
    \end{axis}
    \end{tikzpicture}
    \caption{\textit{Impact of Strategic Planning Agent on Audit Efficiency}. The planning-driven mode detects critical vulnerabilities significantly faster than the ablated baseline. Horizontal grid lines indicate 10-minute intervals for clearer scale reference.}
    \label{fig:efficiency}
\end{figure}

\textbf{Note on Comparisons:} Prior work on smart contract repair (e.g., SmartFix achieving $\approx$94.8\% for vulnerability patching) addresses a different task: repairing production code vulnerabilities. PFIR addresses generated test code failures (compilation errors, missing imports, incorrect assertions). Direct comparison is not meaningful, we report PFIR's 94\% success rate as evidence of effectiveness for the test repair task, with 64\% deterministic fixes reducing LLM costs.

\subsection{RQ4: How do MAS components (protocols, autonomy, self-healing) contribute to system robustness?}

\paragraph{\textbf{Experimental Setup.}}
We conducted systematic ablation studies to isolate MAS contributions across 20 audit sessions, injecting controlled failures (tool crashes, network timeouts, LLM API failures). We compare five configurations: 
\begin{itemize}
    \item Full SPEAR: complete MAS framework
    \item No Protocols: agents operate autonomously but use direct method calls instead of protocols (no Contract Net, plan negotiation, or auctions)
    \item No Autonomy: agents follow fixed policies without local decision-making (centralized controller with same logic)
    \item No Self-Healing: repair agent disabled, failures require manual intervention
    \item Rigid Pipeline: sequential execution, no coordination, no autonomy, no self-healing. We measure recovery time, resource efficiency, coordination overhead, and resilience to failure.
\end{itemize}

\paragraph{\textbf{Results.}}
Table~\ref{tab:mas_properties} summarizes the ablation study isolating coordination protocols, agent autonomy, and self-healing behavior under injected failures.

\textbf{Protocol Contribution}: Comparison of Full SPEAR vs. No Protocols shows that protocols enable 1.8× faster recovery (2.3 min vs. 4.1 min, p < 0.01) and 15\% reduction in LLM invocations (1.5 vs. 1.76, p < 0.05). Protocols enable distributed decision-making with partial observability, allowing agents to negotiate priorities without a central arbitrator.

\textbf{Autonomy Contribution}: Comparing Full SPEAR vs. No Autonomy shows that autonomy enables 2.1× faster recovery (2.3 min vs. 4.8 min, p < 0.01) and 18\% reduction in LLM invocations (1.5 vs. 1.83, p < 0.05). Local decision-making allows agents to adapt to failures without waiting for the central controller.

\textbf{Self-Healing Contribution}: Comparing Full SPEAR vs. No Self-Healing shows that self-healing enables 3.2× faster recovery (2.3 min vs. 7.4 min, p < 0.001) and 31\% reduction in LLM invocations (1.5 vs. 2.17, p < 0.001). Autonomous repair prevents cascading failures.

\textbf{Coordination Overhead}: Full SPEAR has 4.2\% overhead (std: 1.1\%), with a mean of 47 messages per audit. Protocol breakdown: Plan Negotiation (1.8\%), Contract Net (1.5\%), Resource Auction (0.9\%). The overhead scales linearly with the number of agents: $O(|\mathcal{A}|)$ messages per protocol instance.

\begin{table}[ht]
  \centering
  \caption{Ablation study of coordination protocols, agent autonomy, and self-healing under injected failures (20 audit runs).}
  \label{tab:mas_properties}
    \resizebox{\columnwidth}{!}{%
    \setlength{\tabcolsep}{3pt}
  \begin{tabular}{lccccc}
    \toprule
    \textbf{Metric} & \textbf{Full SPEAR} & \textbf{No Protocols} & \textbf{No Autonomy} & \textbf{No Self-Healing} & \textbf{Rigid Pipeline} \\
    \midrule
    Recovery Time (min) & \textbf{2.3 ± 0.8} & 4.1 ± 1.2 & 4.8 ± 1.5 & 7.4 ± 2.3 & 8.7 ± 2.1 \\
    LLM Invocations/Repair & \textbf{1.5 ± 0.4} & 1.76 ± 0.5 & 1.83 ± 0.5 & 2.17 ± 0.6 & 1.95 ± 0.5 \\
    Negotiation Success Rate & \textbf{94\%} & N/A & N/A & 94\% & N/A \\
    Conflict Frequency/Audit & \textbf{1.2 ± 0.6} & 2.8 ± 1.1 & 0 & 1.2 ± 0.6 & 0 \\
    Messages/Audit & \textbf{47 ± 8} & 12 ± 3 & 0 & 47 ± 8 & 0 \\
    Coordination Overhead & \textbf{4.2\% ± 1.1\%} & 1.8\% ± 0.5\% & 0\% & 4.2\% ± 1.1\% & 0\% \\
    \bottomrule
  \end{tabular}
  }
\end{table}

The ablation results indicate distinct roles for coordination protocols, agent autonomy,
and self-healing behavior under injected failures.

\subsection{Threats to Validity}

\textit{Internal Validity:} Controlled failure injection (RQ4) may not capture all real-world failure modes. Ablation comparisons assume implementations are equivalent except for the ablated component, and subtle differences could confound the results. We mitigated this by using identical codebases with feature flags.

\textit{External Validity:} Evaluation uses a DVD benchmark (synthetic) and one DeFi protocol (47 contracts). The results may not be generalized to all smart contract projects. The completeness assumption of the tool (A2 in the proof of liveness) is strong, in practice, tools miss vulnerabilities, resulting in a probabilistic rather than a guaranteed liveness. The sample sizes (15 challenges, 20 audit sessions) are modest, and larger-scale evaluation is future work. 
Although the domain is smart contract auditing, the coordination patterns studied are not domain-specific and may apply to other long-running tool orchestration tasks, we do not claim generality beyond similar failure-prone workflows.

\textit{Construct Validity:} F1-Score and TFCV are standard metrics but may not capture all dimensions of audit quality (e.g., report clarity, false positive cost). The PFIR success rate measures compilation/execution, not semantic correctness of the generated tests.

\subsection{Reproducibility}
\label{sec:reproducibility}

\textbf{Datasets:} Damn Vulnerable DeFi benchmark (15 challenges), 500 security objectives from GitHub Solidity contracts, DeFi protocols (Uniswap V2, Compound V2, Aave V2). \textit{Hyperparameters:} Risk weights $\alpha=0.5$, $\beta=0.3$, $\gamma=0.2$, plan revision threshold $\theta=0.15$, PFIR max attempts=5, \textit{Sensitivity Analysis:} Risk weights $\pm 0.1$ variation changes TFCV by $<$8\%, plan revision threshold $\theta \in [0.10, 0.20]$ balances responsiveness/stability, variation of confidence parameters $\pm 0.1$ changes the repair success rate by $<$3\% (robust). \textit{LLM:} Claude Sonnet 4.5 via Anthropic API, average cost \$2.30 per audit. \textit{Environment:} Ubuntu 22.04, Python 3.10, Foundry 0.2.0, Slither 0.10.0, Mythril 0.24.0. Random seed: 42 for reproducibility. 

\section{Conclusion and Future Work}
\label{sec:conclusion}

This paper presented SPEAR as an engineering case study of multi-agent coordination
applied to autonomous smart contract auditing. The results suggest that explicit
coordination protocols, local agent autonomy, and self-healing policies simplify
recovery and adaptation in long-running, failure-prone workflows. This work illustrates how established coordination mechanisms can be composed to support robustness and resource-aware behavior in
a realistic application domain.

\textbf{Limitations:} Tool completeness (A2) is a strong assumption, real tools miss vulnerabilities. The evaluation scale is modest. The Coordinator Agent ($A_{Coord}$) is a potential single point of failure.

\textbf{Future work:} Policy learning through RL, expanded tool integration, human-in-the-loop interfaces, and distributed coordination without central coordinator.

 \bibliographystyle{splncs04}
\bibliography{b}

@article{wu2023smartllmsentry, 
    title={SmartLLMSentry: A Comprehensive LLM Based Smart Contract Vulnerability Detection Framework}, 
    journal={Journal of Metaverse}, 
    volume={4}, 
    pages={126–137}, 
    year={2024}, 
    DOI={10.57019/jmv.1489060}, 
    author={Zaazaa, Oualid and El Bakkali, Hanan}, 
    number={2}, 
    publisher={İzmir Academy Association}
}

@inproceedings{grieco2020echidna,
author = {Grieco, Gustavo and Song, Will and Cygan, Artur and Feist, Josselin and Groce, Alex},
title = {Echidna: effective, usable, and fast fuzzing for smart contracts},
year = {2020},
isbn = {9781450380089},
publisher = {Association for Computing Machinery},
address = {New York, NY, USA},
url = {https://doi.org/10.1145/3395363.3404366},
doi = {10.1145/3395363.3404366},
booktitle = {Proceedings of the 29th ACM SIGSOFT International Symposium on Software Testing and Analysis},
pages = {557–560},
numpages = {4},
location = {Virtual Event, USA},
series = {ISSTA 2020}
}

@inproceedings{luu2016making,
author = {Luu, Loi and Chu, Duc-Hiep and Olickel, Hrishi and Saxena, Prateek and Hobor, Aquinas},
title = {Making Smart Contracts Smarter},
year = {2016},
isbn = {9781450341394},
publisher = {Association for Computing Machinery},
address = {New York, NY, USA},
url = {https://doi.org/10.1145/2976749.2978309},
doi = {10.1145/2976749.2978309},
booktitle = {Proceedings of the 2016 ACM SIGSAC Conference on Computer and Communications Security},
pages = {254--269},
numpages = {16},
location = {Vienna, Austria},
series = {CCS '16}
}

@article{yu2020screpair,
author = {Yu, Xiao Liang and Al-Bataineh, Omar and Lo, David and Roychoudhury, Abhik},
title = {Smart Contract Repair},
year = {2020},
issue_date = {October 2020},
publisher = {Association for Computing Machinery},
address = {New York, NY, USA},
volume = {29},
number = {4},
issn = {1049-331X},
url = {https://doi.org/10.1145/3402450},
doi = {10.1145/3402450},
journal = {ACM Trans. Softw. Eng. Methodol.},
month = sep,
articleno = {27},
numpages = {32}
}

@misc{sun2024large,
      title={Generative Large Language Model usage in Smart Contract Vulnerability Detection}, 
      author={Peter Ince and Jiangshan Yu and Joseph K. Liu and Xiaoning Du},
      year={2025},
      eprint={2504.04685},
      archivePrefix={arXiv},
      primaryClass={cs.CR},
      url={https://arxiv.org/abs/2504.04685}, 
}

@misc{certora-prover,
  author = {{Certora}},
  title  = {Certora Prover Documentation},
  year   = {2026},
  howpublished = {\url{https://docs.certora.com/}},
  note   = {Accessed: April 6, 2026}
}

@inproceedings{feist2019slither,
author = {Feist, Josselin and Greico, Gustavo and Groce, Alex},
title = {Slither: a static analysis framework for smart contracts},
year = {2019},
publisher = {IEEE Press},
url = {https://doi.org/10.1109/WETSEB.2019.00008},
doi = {10.1109/WETSEB.2019.00008},
booktitle = {Proceedings of the 2nd International Workshop on Emerging Trends in Software Engineering for Blockchain},
pages = {8–15},
numpages = {8},
location = {Montreal, Quebec, Canada},
series = {WETSEB '19}
}

@misc{mueller2021mythril,
  author       = {ConsenSys Diligence},
  title        = {Mythril - a symbolic-execution tool for EVM bytecode},
  howpublished = {\url{https://github.com/ConsenSysDiligence/mythril}},
  month        = mar,
  year         = 2024,
  note         = {Accessed: November 6, 2025}
}

@inproceedings{zhang2024gptscan, 
   series={ICSE ’24},
   title={GPTScan: Detecting Logic Vulnerabilities in Smart Contracts by Combining GPT with Program Analysis},
   url={http://dx.doi.org/10.1145/3597503.3639117},
   DOI={10.1145/3597503.3639117},
   booktitle={Proceedings of the IEEE/ACM 46th International Conference on Software Engineering},
   publisher={ACM},
   author={Sun, Yuqiang and Wu, Daoyuan and Xue, Yue and Liu, Han and Wang, Haijun and Xu, Zhengzi and Xie, Xiaofei and Liu, Yang},
   year={2024},
   month=apr, pages={1–13},
   collection={ICSE ’24} }

@misc{liu2025adaptive,
      title={Adaptive Plan-Execute Framework for Smart Contract Security Auditing}, 
      author={Zhiyuan Wei and Jing Sun and Zijian Zhang and Zhe Hou and Zixiao Zhao},
      year={2025},
      eprint={2505.15242},
      archivePrefix={arXiv},
      primaryClass={cs.CR},
      url={https://arxiv.org/abs/2505.15242}
}

@inproceedings{hildenbrandt2018k,
author = {Chen, Xiaohong and Ro\c{s}u, Grigore},
title = {—A Semantic Framework for Programming Languages and Formal Analysis},
year = {2019},
isbn = {978-3-030-55088-2},
publisher = {Springer-Verlag},
address = {Berlin, Heidelberg},
url = {https://doi.org/10.1007/978-3-030-55089-9\_4},
doi = {10.1007/978-3-030-55089-9_4},
booktitle = {Engineering Trustworthy Software Systems: 5th International School, SETSS 2019, Chongqing, China, April 21–27, 2019, Tutorial Lectures},
pages = {122–158},
numpages = {37},
location = {Chongqing, China}
}

@inproceedings{hildenbrandt2018kevm,
author = {Hildenbrandt, Everett and Saxena, Manasvi and Rodrigues, Nishant and Zhu, Xiaoran and Daian, Philip and Guth, Dwight and Moore, Brandon and Park, Daejun and Zhang, Yi and Stefanescu, Andrei and Rosu, Grigore},
title = {KEVM: A Complete Formal Semantics of the Ethereum Virtual Machine},
year = {2018},
publisher = {IEEE},
url = {https://doi.org/10.1109/CSF.2018.00022},
doi = {10.1109/CSF.2018.00022},
booktitle = {2018 IEEE 31st Computer Security Foundations Symposium (CSF)},
pages = {204--217},
location = {Oxford, United Kingdom}
}

@inproceedings{bhargavan2016formal,
author = {Bhargavan, Karthikeyan and Delignat-Lavaud, Antoine and Fournet, C\'{e}dric and Gollamudi, Anitha and Gonthier, Georges and Kobeissi, Nadim and Kulatova, Natalia and Rastogi, Aseem and Sibut-Pinote, Thomas and Swamy, Nikhil and Zanella-B\'{e}guelin, Santiago},
title = {Formal Verification of Smart Contracts: Short Paper},
year = {2016},
isbn = {9781450345743},
publisher = {Association for Computing Machinery},
address = {New York, NY, USA},
url = {https://doi.org/10.1145/2993600.2993611},
doi = {10.1145/2993600.2993611},
booktitle = {Proceedings of the 2016 ACM Workshop on Programming Languages and Analysis for Security},
pages = {91–96},
numpages = {6},
location = {Vienna, Austria},
series = {PLAS '16}
}

@book{wooldridge2009introduction,
author = {Woolridge, Michael and Wooldridge, Michael J.},
title = {Introduction to Multiagent Systems},
year = {2001},
isbn = {047149691X},
publisher = {John Wiley \& Sons, Inc.},
address = {USA}
}

@INPROCEEDINGS{mariani2008multi,
  author={El Yamany, Hany F. and M. Capretz, Miriam A. and Capretz, Luiz F.},
  booktitle={30th Annual International Computer Software and Applications Conference (COMPSAC'06)}, 
  title={A Multi-Agent Framework for Testing Distributed Systems}, 
  year={2006},
  volume={2},
  number={},
  pages={151-156},
  doi={10.1109/COMPSAC.2006.98}}

@article{haeberle2016grand,
    author = {Mehar, Izhar and Shier, Charles and Giambattista, Alana and Gong, Elgar and Fletcher, Gabrielle and Sanayhie, Ryan and Kim, Henry and Laskowski, Marek},
    year = {2019},
    month = {01},
    pages = {19-32},
    title = {Understanding a Revolutionary and Flawed Grand Experiment in Blockchain: The DAO Attack},
    volume = {21},
    journal = {Journal of Cases on Information Technology (JCIT)},
    doi = {https://doi.org/10.4018/JCIT.2019010102}
}

@inproceedings{werner2021sok,
author = {Werner, Sam and Perez, Daniel and Gudgeon, Lewis and Klages-Mundt, Ariah and Harz, Dominik and Knottenbelt, William},
title = {SoK: Decentralized Finance (DeFi)},
year = {2023},
isbn = {9781450398619},
publisher = {Association for Computing Machinery},
address = {New York, NY, USA},
url = {https://doi.org/10.1145/3558535.3559780},
doi = {10.1145/3558535.3559780},
booktitle = {Proceedings of the 4th ACM Conference on Advances in Financial Technologies},
pages = {30–46},
numpages = {17},
location = {Cambridge, MA, USA},
series = {AFT '22}
}

@article{rao1995bdi,
    author = {Rao, Anand S. and Georgeff, Michael P.},
    title = {BDI Agents: From Theory to Practice},
    journal = {Proceedings of the First International Conference on Multi-Agent Systems (ICMAS-95)},
    year = {1995},
    pages = {312--319},
    publisher = {MIT Press}
}

@article{smith1980contract,
    author = {Smith, Reid G.},
    title = {The Contract Net Protocol: High-Level Communication and Control in a Distributed Problem Solver},
    journal = {IEEE Transactions on Computers},
    volume = {C-29},
    number = {12},
    year = {1980},
    pages = {1104--1113},
    doi = {10.1109/TC.1980.1675516}
}

@techreport{fipa2000acl,
    author = {{Foundation for Intelligent Physical Agents}},
    title = {FIPA Agent Communication Language Specification},
    institution = {FIPA},
    year = {2000},
    type = {Standard},
    number = {SC00061J},
    url = {http://www.fipa.org/specs/fipa00061/}
}

@InProceedings{bellifemine2007jade,
author="Bellifemine, Fabio
and Poggi, Agostino
and Rimassa, Giovanni",
editor="Castelfranchi, Cristiano
and Lesp{\'e}rance, Yves",
title="Developing Multi-agent Systems with JADE",
booktitle="Intelligent Agents VII Agent Theories Architectures and Languages",
year="2001",
publisher="Springer Berlin Heidelberg",
address="Berlin, Heidelberg",
pages="89--103",
isbn="978-3-540-44631-6"
}

@inproceedings{rao1991modeling,
author = {Rao, Anand S. and Georgeff, Michael P.},
title = {Modeling rational agents within a BDI-architecture},
year = {1991},
isbn = {1558601651},
publisher = {Morgan Kaufmann Publishers Inc.},
address = {San Francisco, CA, USA},
booktitle = {Proceedings of the Second International Conference on Principles of Knowledge Representation and Reasoning},
pages = {473–484},
numpages = {12},
location = {Cambridge, MA, USA},
series = {KR'91}
}

@article{kephart2003vision,
    author = {Kephart, Jeffrey O. and Chess, David M.},
    title = {The Vision of Autonomic Computing},
    journal = {Computer},
    volume = {36},
    number = {1},
    year = {2003},
    pages = {41--50},
    publisher = {IEEE},
    doi = {10.1109/MC.2003.1160055}
}

@book{weyns2020introduction,
    author = {Weyns, Danny},
    title = {An Introduction to Self-Adaptive Systems: A Contemporary Software Engineering Perspective},
    year = {2020},
    publisher = {John Wiley \& Sons},
    isbn = {978-1-119-57494-4}
}

@ARTICLE{le2011genprog,
  author={Le Goues, Claire and Nguyen, ThanhVu and Forrest, Stephanie and Weimer, Westley},
  journal={IEEE Transactions on Software Engineering}, 
  title={GenProg: A Generic Method for Automatic Software Repair}, 
  year={2012},
  volume={38},
  number={1},
  pages={54-72},
  doi={10.1109/TSE.2011.104}}

@inproceedings{long2016automatic,
    author = {Long, Fan and Rinard, Martin},
    title = {Automatic Patch Generation by Learning Correct Code},
    booktitle = {Proceedings of the 43rd ACM SIGPLAN-SIGACT Symposium on Principles of Programming Languages},
    year = {2016},
    pages = {298--312},
    doi = {10.1145/2837614.2837617}
}

@inproceedings{feng2020codebert,
    author = {Feng, Zhangyin and Guo, Daya and Tang, Duyu and Duan, Nan and Feng, Xiaocheng and Gong, Ming and Shou, Linjun and Qin, Bing and Liu, Ting and Jiang, Daxin and Zhou, Ming},
    title = {CodeBERT: A Pre-Trained Model for Programming and Natural Languages},
    booktitle = {Findings of the Association for Computational Linguistics: EMNLP 2020},
    year = {2020},
    pages = {1536--1547},
    doi = {10.18653/v1/2020.findings-emnlp.139}
}

@article{sandholm1999distributed,
    author = {Sandholm, Tuomas W.},
    title = {Distributed Rational Decision Making},
    journal = {Multiagent Systems: A Modern Approach to Distributed Artificial Intelligence},
    year = {1999},
    pages = {201--258},
    publisher = {MIT Press}
}

@article{durfee1999distributed,
    author = {Durfee, Edmund H.},
    title = {Distributed Problem Solving and Planning},
    journal = {Multiagent Systems: A Modern Approach to Distributed Artificial Intelligence},
    year = {2001},
    pages = {121--164},
    publisher = {MIT Press}
}

@article{agm1985revision,
    author = {Alchourr\'{o}n, Carlos E. and G\"{a}rdenfors, Peter and Makinson, David},
    title = {On the Logic of Theory Change: Partial Meet Contraction and Revision Functions},
    journal = {Journal of Symbolic Logic},
    volume = {50},
    number = {2},
    year = {1985},
    pages = {510--530},
    doi = {10.2307/2274239}
}
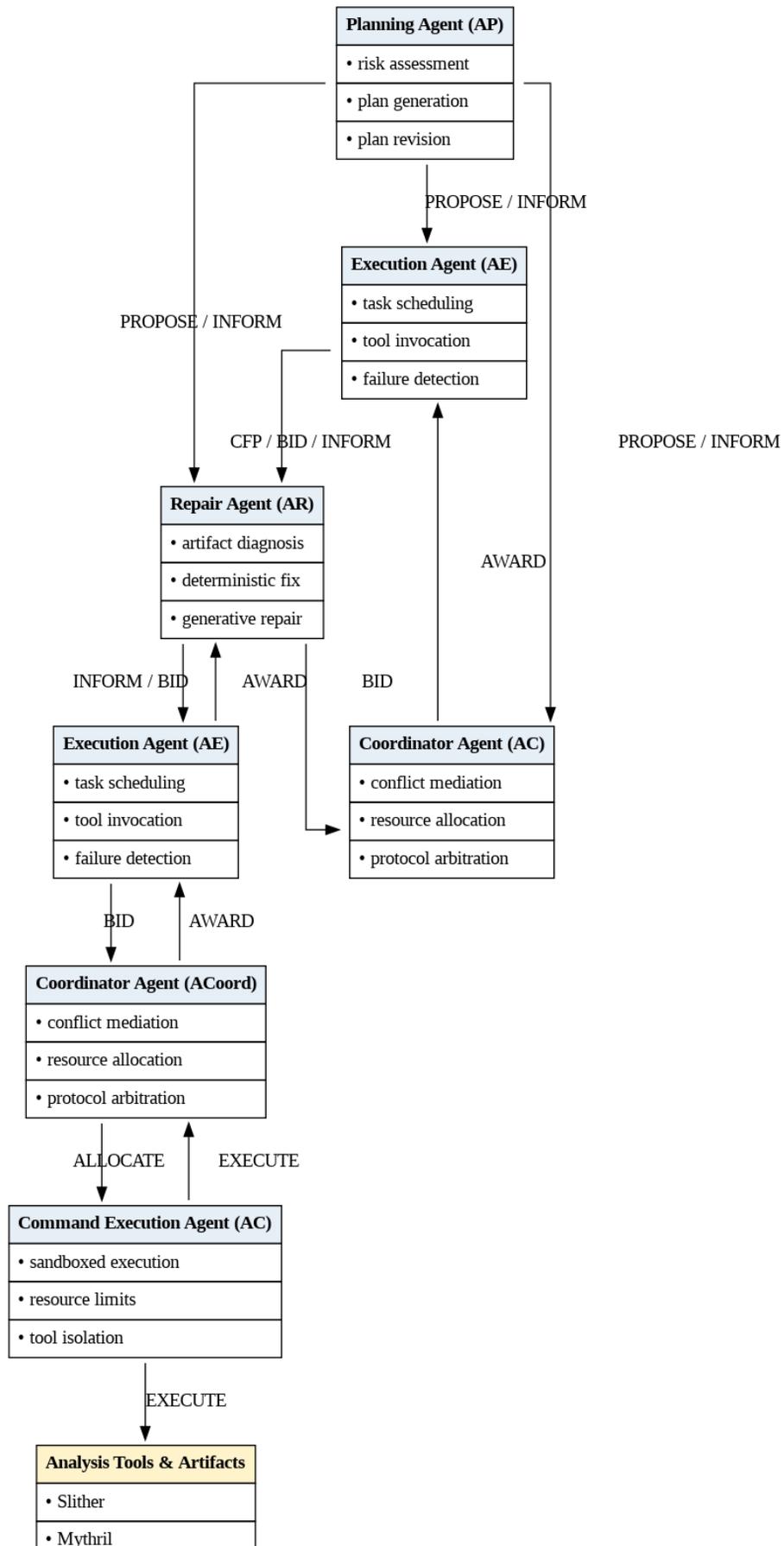
\end{document}